\pdfoutput=1

\documentclass[11pt]{article}

\usepackage[]{acl}

\usepackage{times}
\usepackage{latexsym}

\usepackage[T1]{fontenc}

\usepackage[utf8]{inputenc}

\usepackage{tabularx}

\usepackage{arydshln}
\usepackage{amsmath}

\usepackage{amsfonts}
\usepackage{stmaryrd}

\usepackage{xcolor}

\newcolumntype{Y}{>{\centering\arraybackslash}X}

\usepackage{multirow}
\usepackage{graphicx}
\usepackage{tabularx}

\usepackage{arydshln}

\usepackage{lscape}

\usepackage{microtype}

\title{Medical Coding with Biomedical Transformer Ensembles and Zero/Few-shot Learning}

\author{
  Angelo Ziletti*\\
Bayer AG\\
  {\small\texttt{angelo.ziletti@bayer.com} }
\\\And
Alan Akbik \\
Humboldt University \\
\\\And
Christoph Berns \\
Bayer AG\\
\AND
Thomas Herold \\
areto consulting Gmbh\\
\\\And
Marion Legler \\
Bayer AG\\
\\\And
Martina Viell \\
Bayer AG\\
\\}

\begin{document}
\maketitle
\begin{abstract}

Medical coding (MC) is an essential pre-requisite for reliable data retrieval and reporting. Given a free-text \textit{reported term} (RT) such as ``pain of right thigh to the knee'', the task is to identify the matching \textit{lowest-level term} (LLT) --in this case ``unilateral leg pain''-- from a very large and continuously growing repository of standardized medical terms. However, automating this task is challenging due to a large number of LLT codes (as of writing over $80\,000$), limited availability of training data for long tail/emerging classes, and the general high accuracy demands of the medical domain.
With this paper, we introduce the MC task, discuss its challenges, and present a novel approach called \textsc{xTARS} that combines traditional BERT-based classification with a recent zero/few-shot learning approach (\textsc{TARS}). We present extensive experiments that show that our combined approach outperforms strong baselines, especially in the few-shot regime. 
The approach is developed and deployed at Bayer, live since November 2021. As we believe our approach potentially promising beyond MC, and to ensure reproducibility, we release the code to the research community. 



\end{abstract}

\section{Introduction} \label{sec:Intro}
Medical coding (MC) is the process of classifying textual descriptions of medical events into standardized alphanumerical terms and codes. An example textual description
is ``pain of right thigh to the knee'' that would need to be classified as an instance of ``unilateral leg pain'' in the MedDRA~\citep{meddra} ontology (see Table~\ref{tab:ex-data} for more examples).

%
%
MC allows the consistent documentation of medical records, enabling the analysis of clinical trials, for example facilitating safety data retrieval or detection of adverse drug reactions.
Medical codes are also used by health plan, medical billing, and health care providers to make decisions for example about prior authorization requests and claims, impacting how much a patient will pay for medical care in some countries. At Bayer, around $55\,000$ terms per month need to be manually coded via a ``four-eye concept'' (proposing/accepting) by highly specialized medical coders, a costly process we seek to (partially) automate. 

\begin{table*}[thb]
\small
\centering 
\begin{tabular}{l l l } 
\begin{tabularx}{0.95\textwidth}{X|X|X}
\hline\hline 
Reported term (RT) & Lowest level term (LLT) name  &  Preferred term (PT) name \\
\hline 
on and off lethargy & lethargy & lethargy \\
\hdashline[0.5pt/5pt]
scattered indeterminate subcentimeter pulmonary nodules & lung nodule & pulmonary mass \\
\hdashline[0.5pt/5pt]
ckd-unknown etiology & chronic kidney disease & chronic kidney disease \\
\hdashline[0.5pt/5pt]
elective left fem-pop bypass graft & femoropopliteal artery bypass & peripheral artery bypass \\
\hdashline[0.5pt/5pt]
osteomyelitis of the left metatarsal & osteomyelitis of the foot & osteomyelitis \\
\hdashline[0.5pt/5pt]
right foot second toe gangrene & gangrene toe & gangrene \\
\hdashline[0.5pt/5pt]
worsenin renal function & renal function aggravated & renal impairment \\
\hdashline[0.5pt/5pt]
oliguric acute kidney injury/ckd & acute oliguric renal failure & acute kidney injury \\
\hdashline[0.5pt/5pt]
urinary tract infection [enterobacter cloacae] & urinary tract infection bacterial & urinary tract infection bacterial \\
\hdashline[0.5pt/5pt]
skin defect [no split] & skin disorder& skin disorder \\
\hdashline[0.5pt/5pt]
haemangioma th12 & spinal haemangioma & haemangioma of bone \\
\hdashline[0.5pt/5pt]
pain of right thigh to the knee & unilateral leg pain & pain in extremity \\
\hline\hline 
\end{tabularx}
\end{tabular} 
\caption{Sample medical coding data. \textit{Reported terms} (RT) are short, free-form medical event descriptions that need to be classified into the most suitable \textit{lowest level term} (LLT), from a total of over $80\,000$ standardized LLTs. Each LLT belongs to a \textit{preferred term} (PT), i.e.~a less granular category of classes. For instance ``cdk-unknown etiology'' should be normalized to ``chronic kidney disease''. \label{tab:ex-data}} 
\end{table*}

\noindent 
\textbf{Problem Statement.} However, automating MC faces several challenges: First, the number of target classes is very large and continuously growing, with over $80\,000$ as of writing. Second, available training data is limited and imbalanced, with few training examples in particular available for long tail and emerging classes. Third, language is non-canonical and domain-specific, with frequent misspellings, non-standard abbreviations, irrelevant text, and specialized vocabulary. Forth, as is standard in the medical domain, very high accuracy requirements apply (see Section~\ref{sec:data-sources}).

Conceptually, this task may be phrased in two ways: (1) as a standard large-scale multiclass classification task that takes as input a reported term and outputs a distribution over all classes~\cite{chalkidis-etal-2020-empirical}, or (2) as a matching task that takes as input both a reported term and a candidate class label and makes a binary prediction whether the class matches the term. The latter allows the model to leverage additional information conveyed by the natural language class labels (i.e.~allowing the model to learn that the semantics of the class description ``unilateral leg pain'' and the text ``pain of right thigh to the knee'' overlap) and has thus been shown work well in few-shot learning settings~\cite{halder2020}. It suffers however from scalability issues that prevent practical application to large-scale classification problems.

\noindent 
\textbf{Contributions.} With this paper, we present a novel approach that addresses the above challenges by integrating a classic BERT-based classification approach~\cite{devlin-etal-2019-bert} into a recently proposed few-shot classification approach~\cite{halder2020}. The main idea is to leverage a standard classifier to predict a set of candidate labels which are then separately evaluated by the few-shot learner. We argue that this architecture allows both components to leverage their respective strengths. 
To summarize, our contributions are as follows: 

\begin{itemize}
\item We present and discuss the MC task, and discuss its challenges in particular with regards to industry application. 
\item We present a novel and straightforward approach called \textsc{xTARS}(eXtreme Task-Aware Representation of Sentences) to address this task by combining strengths of large-scale classification and few-shot learning.
\item We conduct an extensive experimental evaluation that shows that our proposed approach outperforms very strong baselines. We also evaluate ensemble learning setups and discus results in different confidence brackets. 
\item Since we believe this approach to be useful beyond the task of MC, and to ensure reproducibility, we make available our implementation to the research community.\footnote{\url{https://github.com/Bayer-Group/xtars-naacl2022}}
\end{itemize}

The presented approach is deployed since November 2021 at Bayer and is used to generate coding proposals for all clinical trial studies running at the time of writing. 

\section{Task and Data Sources} \label{sec:data-sources}
\begin{table*}[!th]
\small
\centering
\begin{tabularx}{\textwidth}{c | c | c | Y | Y }
\hline
\hline
Data & \# of samples & \# of classes & samples from classes with $\leq$10 samples (\%, label cardinality) & \ samples from classes with $\leq$5 samples (\%, label cardinality)\\
\hline
All & $293\,645$& $26\,893$ &21.0\% \hspace{0.3em} 4.96 & 12.2\% \hspace{0.3em} 2.81\\
Train+Val & $285\,070$  & $26\,692$& 21.1\% \hspace{0.3em} 4.95& 12.3\% \hspace{0.3em} 2.81\\
Test (all) & $8\,575$& $4\,436$&  18.0\% \hspace{0.3em} 5.06& 10.0\% \hspace{0.3em} 2.80\\
Test (top-80\%) & $6\,860$  & $3\,495$&12.7\% \hspace{0.3em} 5.84 & 5.5\% \hspace{0.3em} 3.11\\
Test (btm-50\%) &$4\,287$  & $3\,126$&29.5\% \hspace{0.3em} 4.71 & 18.0\% \hspace{0.3em} 2.72\\
Test (btm-25\%) &$1\,715$  & $1\,455$& 39.1\% \hspace{0.3em} 4.05& 28.0\% \hspace{0.3em} 2.55\\
\hline
\hline
\end{tabularx}
\caption{\label{dataset-table} Summary statistics of the medical coding dataset comprising coded and autocoded data, and company synonyms. Augmented data 
is excluded. Uncertainty from one PubMedBERT model is used for test set splits.}
\end{table*}
The MC input is the textual description of a medical event, known as \emph{reported term} (RT).
The goal of MC is to associate a given RT to the most appropriate term from a given ontology. 

\subsection{MedDRA as target ontology}

We leverage the MedDRA \citep{meddra} ontology, which is organized in a multi-level hierarchy with coarse- and fine-grained classes. The more fine-grained level of the hierarchy is the \emph{lowest level term} (LLT), of which approximately $80\,000$ distinct classes exist as of writing. A more coarse-grained level is the \emph{preferred term} (PT), of which approximately $26\,000$ currently exist in MedDRA. Table~\ref{tab:ex-data} shows a number of examples for RTs and their corresponding MedDRA LLT and PT names. 

MedDRA undergoes frequent releases that include changes to the number of classes or their definitions. As we are required to always use the most current version of MedDRA, our approach needs to be robust with regards to such changes.

\subsection{Training data}
We use a number of proprietary data sources to train and evaluate our proposed approach. The first is \textit{coded data}, which are RTs manually linked to LLTs by human experts. The second is \textit{autocoded data} where a simple rule-based system automatically linked those RTs which either are or contain an LLT verbatim. The system has high precision but low recall, with the majority of samples ($\sim$55\%) out of autocoder scope, and passed to humans for manual coding.
In addition, we use a dataset of \textit{company synonyms} consisting of pairs of medical text descriptions and corresponding LLT. These synonyms are created and maintained by the company MC department. These synonyms define concepts that are more specific than LLTs.

\noindent 
\textbf{Final training dataset (Table \ref{dataset-table}).} We collect data from these sources for all Bayer active clinical trials as of October 2021. Data processing and augmentation steps are outlined in the Appendix. The final dataset is split into training, validation and test splits. 

Summary statistics are presented in Table \ref{dataset-table}. We observe that in the entire training data set, only $26\,893$ classes are observed, meaning that a significant portion of MedDRA LLT codes have no training data at all. We further note a significant data imbalance: among the observed classes, $21\,187$ (78\%) have less than 10 samples, with roughly 21\% of all samples coming from those classes. We split the test data into three distinct splits that have different uncertainty, as quantified by the predictive entropy (see Section \ref{sec:exp-setup}). Top-80\% indicates the 80\% more certain data from the test set while btm-50\% and btm-25\% contain the least certain 50\% and 25\% of the test set respectively.

\section{Method} 
We frame the MC task as multiclass classification, where each LLT name is treated as a distinct class. Since each LLT belongs to exactly one PT, the PT is then obtained directly from MedDRA.
We therefore disregard the label hierarchy in MedDRA, as this results in a simpler model to train and deploy, and it makes the model less dependent on topological changes of the underlying MedDRA ontology\footnote{Moreover, \citet{chalkidis-etal-2020-empirical} showed that using label hierarchy information in large-scale multilabel classification is on-par or even inferior to transfer learning approaches.}.

\noindent
\textbf{Method overview.} As outlined in Section~\ref{sec:Intro}, our approach builds on and combines two existing approaches: (1) a default large-scale multiclass prediction approach based on BERT, and (2) a few-shot classification approach. In this section, we first discuss these two baseline approaches and their advantages and drawbacks (Sections~\ref{sec:bert} and~\ref{sec:tars}). We then present our \textsc{xTARS} approach in Section~\ref{sec:few-shot-model}. 

\subsection{Baseline 1: Multiclass Classification with BERT Ensembles}
\label{sec:bert}
Our first baseline follows the standard multiclass classification approach based on BERT~ \citep{devlin-etal-2019-bert}: we add a single softmax classifier as ``prediction head'' over the text embedding retrieved from the CLS-token of a pre-trained BERT model. The language model and prediction head are jointly fine-tuned using standard parameters (see Appendix for details) to output a distribution of prediction scores for all classes given a single input text. Such approaches have been applied to large-scale multi-label text classification for biomedical data, showing results comparable to more complex and bespoke approaches~\cite{chalkidis-etal-2020-empirical}.

\noindent
\textbf{Deep ensembles.} 
\label{sec:deep-ensemble}
However, deploying a single model is not advisable for a production setting due to the underspecification problem \citep{amour2020-under}.
Models that perform equally well on their training domain can produce widely different results in their deployment domain, especially under dataset shift. This can result in instabilities 
when models are deployed in a real-world setting. This is particularly problematic for MC, where consistency is an essential requirement.

To mitigate the underspecification problem we use deep ensembles \citep{lakshminarayanan-2017-deep-ensembles}.
The main idea is to train different models which only differ by a random perturbation
 (usually the random seed) 
and then average across these models to increase prediction stability, and possibly performance. As a further advantage, deep ensembles have also shown to produce reliable uncertainty estimates \citep{lakshminarayanan-2017-deep-ensembles, fort2020-deep}, on par with Bayesian deep learning approaches \citep{gal-2016-dropout}.

\subsection{Baseline 2: TARS Few-Shot Classification} 
\label{sec:tars}

Traditional machine learning algorithms do not have access to the natural language definition of the label,
but rather to a discrete representation known as encoding (e.g., one-hot encoding). This representation 
does not preserve any semantic information present in the natural language definition. As a result, the model can learn the class meaning only indirectly, via the samples associated to a given label during learning.

In the few-shot setting, the lack of label semantic is a clear drawback. 
Inspired by natural language inference, Task-Aware Representation of Sentences (\textsc{TARS}, ~\citet{halder2020}) include label semantic by concatenating the input text (e.g., RT) with labels (e.g., LLT name), and then predicting \emph{True} if the label is the correct one, and \emph{False} otherwise (negative classes or samples). They showed that \textsc{TARS} reaches strong results in few-shot and zero-shot settings, but only evaluated on data sets with comparatively small label sets.

\noindent
\textbf{Limitations with regards to medical coding.} For MC, \textsc{TARS} is confronted with severe scalability issues due to the very large number of classes in MedDRA: During prediction, a distinct forward pass through the model needs to be made to separately evaluate each label candidate. This procedure requires $K$ predictions, where $K$ is the number of possible labels. When $K$ is very large (e.g., $K \sim 80\,000$ for MC with MedDRA), calculating predictions becomes computationally prohibitive. 

In addition, the large-scale classification scenario complicates the training procedure. \textsc{TARS} employs a hard-negative sampling technique to sample a set of $neg$ plausible negatives for each labeled data point, sampling with a probability proportional to the cosine similarity between the correct label and the given label. Since the similarity is used as drawing probability for negative labels, in large-scale classification (as in MC), the model will often see negative labels that are ``too easy''. This hinders learning with ultra-fine-grained labels.

\subsection{Proposed Approach: \textsc{xTARS}} 
\label{sec:few-shot-model}

We introduce changes to the \textsc{TARS} algorithm that improve on negative sampling for training, and address the complexity issues during predictions. Our approach leverages a default BERT multiclass classification model (see Section~\ref{sec:bert}) that must first be separately trained for MC. 

\noindent 
\textbf{Sampling hard negatives in large label sets.} To address the above-mentioned issue on sampling difficult negatives in very large label sets, we propose two sampling techniques that we use jointly. The first leverages predictions from the trained BERT classification model. We use its top-5 predictions (or top-4 if the correct class is in the top-5) as negative samples, as these are hard from the point of view of a fully trained standard model.

The second modifies \textsc{TARS}' cosine similarity-based sampling using top-$k$ and softmax rescaling: After computing label similarities, we extract only the top-$k$ similar classes (out of all $K$ classes) to the correct label, and set all others to zero. We choose $k$ to be three times the number of negative samples to be drawn. Finally, we rescale the top-$k$ similarities via temperature-scaled softmax with temperature $T$. 
Low (high) $T$ will result in more peaked (broad) distribution. 
This procedure improves the quality of negative samples (Table \ref{benchmark-table}, cf. \textit{TARS (neg=10)} and \textit{xTARS (neg=10)}), and it is faster (sampling probability vector of $k$  instead of $K$ dimensions, with $k \ll K$) .

\noindent 
\textbf{Limiting candidate labels during prediction.} We address the scalability issues in prediction by first predicting a multiclass distribution with a default BERT model (or deep ensemble). We select the \textit{n} top-scoring predictions to be used as label candidates for \textsc{TARS}. Through experimentation, we found a good value of \textit{n} to be 5. This leads to a four-order of magnitude reduction in computational cost (5 vs $80\,000$) with only a small decrease in accuracy. The obvious drawback is that \textsc{xTARS} cannot predict correctly if the correct label is not in the top-$5$ BERT candidates; however, the number of candidates can be increased so that a target cumulative accuracy is reached.

\begin{table*}[h!t]
\setlength{\tabcolsep}{4pt}
\small
\centering
\begin{tabular}{l X l X l X l X l X l X l X l X l X l }
\hline
\hline
\multirow{2}{4em}{Model} &  
\multicolumn{4}{| c|}{LLT Accuracy [\%]} & \multicolumn{4}{ c |}{PT accuracy [\%]} \\
& All & top-80\% & btm-50\% & btm-25\% & All & top-80\% & btm-50\% & btm-25\% \\

\hline
PubMedBERT (PMB) (single)  & 74.9$_{0.5}$ & 83.9$_{0.3}$& 57.3$_{0.9}$ & 42.1$_{1.4}$ & 88.9$_{0.2}$ &  \textbf{95.8}$_{0.1}$ & 79.8$_{0.3}$ & 66.5$_{0.6}$  \\
BioBERT (BB) (single)  & 74.9$_{0.3}$ & 83.7$_{0.2}$& 56.9$_{0.7}$ & 42.6$_{1.3}$&  88.8$_{0.2}$ & \textbf{95.8}$_{0.2}$ & 79.7$_{0.3}$ & 66.2$_{0.5}$ \\
sciBERT (SB) (single)  & 74.9$_{0.1}$ & 83.5$_{0.1}$& 57.1$_{0.5}$ & 43.8$_{0.7}$& 88.9$_{0.1}$ & \textbf{95.8}$_{0.1}$ & 79.8$_{0.2}$ & 66.7$_{0.4}$ \\
\hdashline[0.5pt/5pt]
TARS (neg=2 cos)  & 64.9$_{1.2}$ & 70.3$_{1.4}$ & 52.2$_{0.9}$ & 44.6$_{0.6}$ & 85.5$_{0.4}$ & 90.7$_{0.5}$ & 78.3$_{0.1}$ & 68.6$_{0.3}$  \\
TARS (neg=10 cos) & 62.6$_{0.8}$ & 67.8$_{0.9}$ & 50.8$_{0.4}$ & 43.8$_{0.7}$ & 85.1$_{0.1}$ & 90.3$_{0.1}$ & 78.0$_{0.3}$ & 68.7$_{0.2}$ \\
\hdashline[0.5pt/5pt]
\textsc{xTARS} (neg=10 cos) & 64.9$_{0.9}$ & 70.4$_{1.1}$ & 51.9$_{0.8}$ & 44.5$_{0.5}$ & 86.1$_{0.4}$ & 91.4$_{0.4}$ & 78.7$_{0.4}$ & 69.3$_{0.3}$  \\
\textsc{xTARS} (neg=top-5) & 76.2$_{0.4}$ & 83.5$_{0.6}$ & 61.7$_{0.4}$ & 51.2$_{0.4}$ & 89.0$_{0.1}$ & 94.7$_{0.2}$ & 81.0$_{0.1}$ & 70.5$_{0.2}$  \\
\textsc{xTARS} (neg=top-5+5 cos)  & 77.3$_{0.2}$ & 84.3$_{0.2}$ & 63.1$_{0.1}$ & 52.4$_{0.4}$ & 89.7$_{0.05}$ & 95.3$_{0.05}$ & 82.1$_{0.1}$ & 71.6$_{0.3}$  \\
\textsc{xTARS} (neg=top-5+10 cos)  & 76.9$_{0.7}$ & 84.0$_{0.7}$ & 62.9$_{0.5}$ & 52.2$_{0.6}$ & 89.8$_{0.3}$ & 95.2$_{0.3}$ & 82.3$_{0.3}$ & 72.3$_{0.1}$  \\
\textsc{xTARS} (neg=top-5+5 cos; $T$=1)   & \textbf{77.5}$_{0.3}$ & \textbf{84.4}$_{0.3}$ & \textbf{63.5}$_{0.2}$ & \textbf{53.5}$_{0.2}$ & \textbf{90.0}$_{0.05}$ & 95.4$_{0.1}$ & \textbf{82.5}$_{0.1}$ & \textbf{72.4}$_{0.1}$  \\
\hline
PMB (3 models) & 77.8 & \textbf{85.7} & 61.2 & 49.2 &90.6& \textbf{96.7}& 82.6& 71.2 \\
BB (3 models)& \textbf{77.9} & 85.2&  \textbf{61.5} &\textbf{51.0} &\textbf{90.7} & 96.5 &\textbf{83.0} & \textbf{72.0} \\
SB (3 models) & \textbf{77.9}& 85.3&  \textbf{61.5}& 50.8& 90.5&96.3& 82.5&  70.9 \\
\hline
PMB+SB+BB (3$\times$3) & 79.7& 86.2&  64.4& 55.4& \textbf{91.7}& 96.7&84.6&  74.9 \\
\textsc{xTARS} (PMB+SB+BB, 3$\times$3) &\textbf{80.4} & \textbf{86.4}&\textbf{64.7}&  \textbf{56.1} &91.6 &\textbf{96.8}& \textbf{84.8}& \textbf{75.5} \\
\hline
PMB+SB+BB (3$\times$5) & 80.1& 86.3& 64.9& 56.8& 92.0& 96.8& 85.1& 76.2\\
\hline
\hline
\end{tabular}
\caption{\label{benchmark-table} Results on the test set. BERT (\textsc{TARS} and \textsc{xTARS}) models are fine-tuned with five (three) different random seeds: average accuracy and standard deviation are reported. Unless otherwise specified, \textsc{xTARS} and TARS are fine-tuned from PubMedBERT. 
If not specified, $T=0.01$. $m$ cos indicates that $m$ negative samples are drawn via the cosine similarity procedure (Sec. \ref{sec:few-shot-model}); top-5 means that the top-5 (or top-4, if the correct class is in the top-5) BERT predictions are used as negative samples. \textsc{xTARS} ensemble is performed with \textsc{xTARS} (neg=top-5+5 cos; $T=1$). In bold the highest accuracy for a fixed number of models (i.e.~1, 3, 9).
}
\end{table*}

\section{Evaluation}

We evaluate our proposed \textsc{xTARS} approach against strong BERT and \textsc{TARS} baselines, both in single-model and ensemble-model setups common to industrial application. Our evaluation tests all approaches in the large-scale multiclass classification scenario of MC, and evaluates the impact of our proposed negative sampling techniques. We also specifically evaluate performance for certain (top-80\%) and uncertain samples (btm-50\% and btm-25\%). The uncertain splits aim at evaluating performance in the few-shot regime (Table \ref{dataset-table})\footnote{We split by uncertainty instead of class frequency because the uncertainty estimation is also available at prediction time, i.e.~during industrial deployment of the model.}.

\subsection{Experimental setup}
\label{sec:exp-setup}
\noindent
\textbf{Language models and ensembles.} We select the top-scoring pre-trained language models for biomedical tasks from the BLURB leaderboard~\citep{gu-2020-pubmedbert}, namely bioBERT~\citep{lee-2019-biobert}, PubMedBERT~\citep{gu-2020-pubmedbert}, and sciBERT~\citep{beltagy-etal-2019-scibert}. For each training run, we train multiple models which differ only in the random seed initialization, and then perform model ensembling via averaging their classification probabilities (see Appendix for more details).

\noindent
\textbf{Estimation of uncertainty.} To obtain principled uncertainty estimates, we utilize the concept of predictive entropy which
%
captures the average amount of information contained in the predictive distribution \citep{gal-2016-dropout}. 
The larger (smaller) the predictive entropy, the more uncertain (certain) the prediction is.
The maximum of the predictive entropy is attained when all prediction probabilities are equal, while it is zero when one probability is equal to one and all the rest are zero.

\subsection{Experimental results}
Results for both LLT and PT are shown in Table \ref{benchmark-table}. As expected, accuracy is higher across all experiments for PT than for LLT, and lower for less certain predictions. In more detail, we make the following observations: 

\noindent 
\textbf{Strong single-model performance for \textsc{xTARS}.} The top 10 rows in Table~\ref{benchmark-table} list the results for single (e.g.~non-ensemble) models. We note that all three BERT models (PubMedBERT, BioBERT, sciBERT) score roughly on-par. \textsc{TARS} underperforms BERT when all samples are considered, especially for LLT; gains are marginal even for very uncertain samples (btm-25\%). Inclusion of more negative samples  (Table \ref{benchmark-table}, cf. \textit{TARS (neg=10 cos)} vs \textit{TARS (neg=2 cos)}) does not improve results. 

Our \textsc{xTARS} models on the other hand for the most part significantly outperform all single-model baselines, reaching an LLT accuracy of 77.5 ($\uparrow$3.6pp from the best single-model BERT) and a PT accuracy of 90 ($\uparrow$1.1pp).

\noindent 
\textbf{Impact of different negative sampling techniques.} Further analyzing these results, we find that \textsc{xTARS} top-$k$ and softmax rescaling sampling improves performance over \textsc{TARS} (Table \ref{benchmark-table}, cf. \textit{TARS (neg=10)} and \textit{\textsc{xTARS} (neg=10))}, but only slightly.
In contrast, inclusion of negative samples from the trained BERT classification model strongly improves performance ($\uparrow$11.3pp in LLT accuracy), making \textsc{xTARS} outperform all BERT models in the single-model setting. This indicates that hard negatives are needed to learn effectively with large label sets.
Combining these two sampling strategies further improves performance.

\noindent 
\textbf{Impact of model uncertainty.} \textsc{xTARS} improves BERT performance overall, with stronger improvements for uncertain samples ($\uparrow$9.7pp for btm-25\%), which is the few-shot regime (Table \ref{dataset-table}). It also reduces the standard deviation for uncertain samples, suggesting an increased stability on random seed initialization w.r.t. BERT and \textsc{TARS} models in the few-shot regime.

\noindent 
\textbf{Ensemble results.} 
Ensembling results in performance gains for all models, with stronger gains observed for BERT over \textsc{xTARS}. Model ensembling improve accuracy overall, including uncertain samples.
Ensembling models from the same and different language model are both beneficial. Gains flatten with more models (cf. 3$\shortrightarrow$9 vs 9$\shortrightarrow$15).

\section{Discussion and practical deployment}

Our experimental evaluation shows that \textsc{xTARS} strongly outperforms all other approaches in the single-model setting, and even slightly outperforms other approaches in the ensemble setting. 

For model deployment, however, other practical considerations need to be taken into account than just overall accuracy. One drawback of \textsc{xTARS} is that it requires a fully trained multiclass classification BERT model before the model can be trained. As our setup is a continuously running system, and both MedDRA and our training data are constantly expanded, we implemented an automated system for retraining models in regular intervals. Here, we decided on the BERT ensemble setup ($3 \times 5$) because of its high accuracy and relatively low complexity.

\noindent
\textbf{Human verification in running system.} After deployment, we perform back-testing to estimate real-world performance: predictions are compared with the label given by the human coder.
For a total of $2\,452$ predictions from the live system, we found a LLT accuracy of 90.9\% with 80.3\% coverage.

From an industry perspective, this accuracy significantly improves coding efficiency. Human coders are presented with a system proposal which they can simply accept in many cases, leaving only a small portion of data points in which coders need to manually search for the best matching LLT code. For the top-80\% most certain samples, nearly all models meet the regulatory requirements of 95\% PT accuracy for MC.  



\section{Related work} \label{sec:related-work}

\noindent 
\textbf{Large-scale text classification.} 
The literature focuses on assigning multiple medical codes to the unstructured portion of electronic health records (patient notes or narratives)
\citep{baumel2018,mullenbach-etal-2018-explainable,rios-kavuluru-2018-shot,shi2017,xie-xing-2018-neural,kim21a}. In our case, however, only text snippets relevant to the coding process (i.e. RT) are gathered during the clinical trial data collection process. Each text snippet must be assigned to a single code. 

\noindent 
\textbf{Zero/few-shot learning.} 
Few-shot learning in NLP has been performed mostly via meta-learning \citep{finn2017}. Meta-learning has been applied for example in machine translation \citep{gu-etal-2018-meta}, sentiment analysis \citep{yu-etal-2018-diverse}, and dialog intent classification \citep{geng-etal-2019-induction}.
However, these approaches cannot perform zero-shot predictions.
\citet{yin-etal-2019-benchmarking} propose to treat zero/few-shot text classification as a textual entailment problem.
The input text acts as premise, and labels are used as hypotheses.
\citet{halder2020} adopt a similar idea.
Literature on zero/few-shot learning in large-scale text classification for biomedical data is scarce \citep{chalkidis-etal-2020-empirical,ijcai2020}. 

\noindent
\textbf{Deployed systems for medical coding.} 
%
MagiCoder is a rule-based system~\citep{zorzi2017, combi2019} to obtain medical codes from pharmacovigilance reports that scans the input text for terms matching the ontology, and votes the best match. They achieved an average precision (recall) of 69\% (70\%) on an adverse drug reaction dataset scraped from social media~\citep{yang2012}.
%

\section{Conclusions}
In this paper, we introduced the MC task and discussed its challenges. 
We outlined a MC system based on biomedical transformers deployed in a production environment, and showed that ensembling improves performance. We introduced \textsc{xTARS}, a zero/few-show learning approach, suitable for classification tasks with very large label sets and long-tailed distribution of labels in data points. The main limitation of \textsc{xTARS} is that it requires a (well-performing) BERT model, thus increasing model complexity. 
We report promising results for \textsc{xTARS} in MC, and release our code to the research community for application to other tasks.

\section*{Acknowledgments}
We thank Adrian Wrobel and Christoph Brieden for the connection with the MC platform, Hanna Viol for providing subject matter expertise, Tanja Bellaire for the validation process, and Stefanie Neumann and Tim Disselhoff for project management.
Alan Akbik is supported by the Deutsche Forschungsgemeinschaft (DFG, German Research Foundation) under Germany’s Excellence Strategy – EXC 2002/1 "Science of Intelligence" (project number 390523135) and the DFG Emmy Noether grant “Eidetic Representations of Natural Language” (project number 448414230).

\bibliography{anthology,custom-rebib}

\appendix

\section{Appendix}

\setcounter{table}{0}
\renewcommand{\thetable}{A\arabic{table}}

\subsection{Data processing and augmentation} \label{sec:data-pp}
\label{sec:appendix}

The proposed algorithm maps \textit{reported terms }(RTs) to lower level terms (LLT) in the MedDRA ontology (see main text). This mapping is already available for the coded and autocoded data, and thus these data sources can be readily used.

\begin{figure}[htb]
\centering
\includegraphics[width=0.45\textwidth]{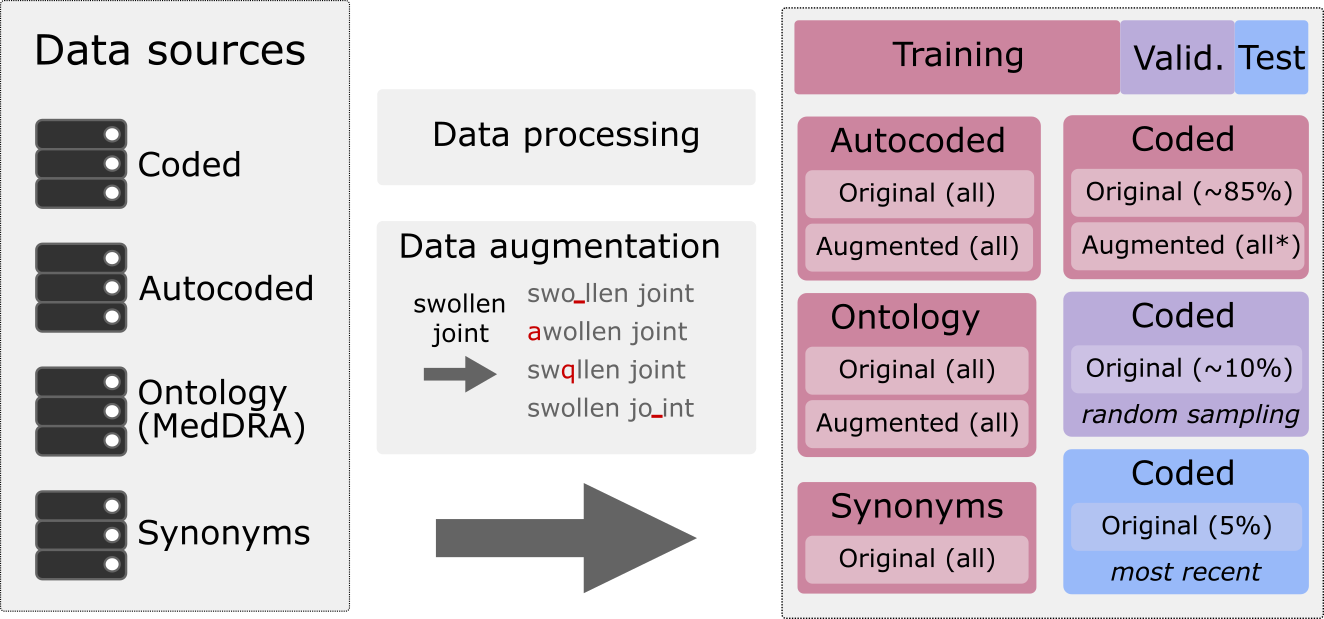}
\caption{From raw data sources to training, validation, and test data via data processing and data augmentation (BERT models).}
\label{fig:data-split-bert}
\end{figure}

\noindent 
\textbf{Data augmentation.}
The dataset presents a large amount of rare classes: more than 66\% of the LLT classes appear less than five times in the data.
To mitigate this problem, for each sample belonging to rare LLTs, we perform data augmentation by generating two simulated samples (one word split and one random character change, see  Fig. \ref{fig:data-split-bert}). We define LLTs as rare if they have less than ten samples. Data augmentation is performed with the \textit{nlpaug} package \citep{ma2019nlpaug}.

\noindent 
\textbf{Data from MedDRA and its augmentation.}
For the MedDRA ontology, we interpret each LLT as RT. For each LLT in the ontology, we generate three RTs: the LLT verbatim, plus two simulated misspelled entries (one word split and one character change), as depicted in Fig. \ref{fig:data-split-bert}. As label, we use the original LLT.
This procedure of adding all possible MedDRA LLT via simulated samples enables the algorithm to make predictions encompassing all possible LLTs, including the ones never encountered in the real (autocoded + coded) data.

\noindent 
\textbf{Company synonyms.}
Each company synonym is interpreted as RT, and the pair RT/LLT is added to the dataset. 
We do not augment synonyms because they are very similar to the corresponding LLT, and the LLTs are already augmented directly from the ontology.

\noindent 
\textbf{Data preprocessing.}
Data preprocessing is minimal: everything is lowercased and only unique RTs are kept; if multiple RT/LLT pairs are present, the most recent pair is kept.

\noindent 
\textbf{Split into training, validation, and test set.}
The last step is to split the data into training, validation, and test set.
The most recent 5\% of coded data is used as test set. From the remaining coded data, a randomly sampled 10\% of the coded data (excluding data augmentation) is used as validation data. 
The training data comprises all autocoded data, all RT originating from the ontology, the company synonyms, and the remaining coded data ($\sim$ 85\%), plus augmented data.
Finally, we remove from the training data the augmented samples if their original sample is included in the validation data. This is done to avoid target leakage due to data augmentation, and enforce the independence of the validation data.

\begin{figure}[htb]
\centering
\includegraphics[width=0.45\textwidth]{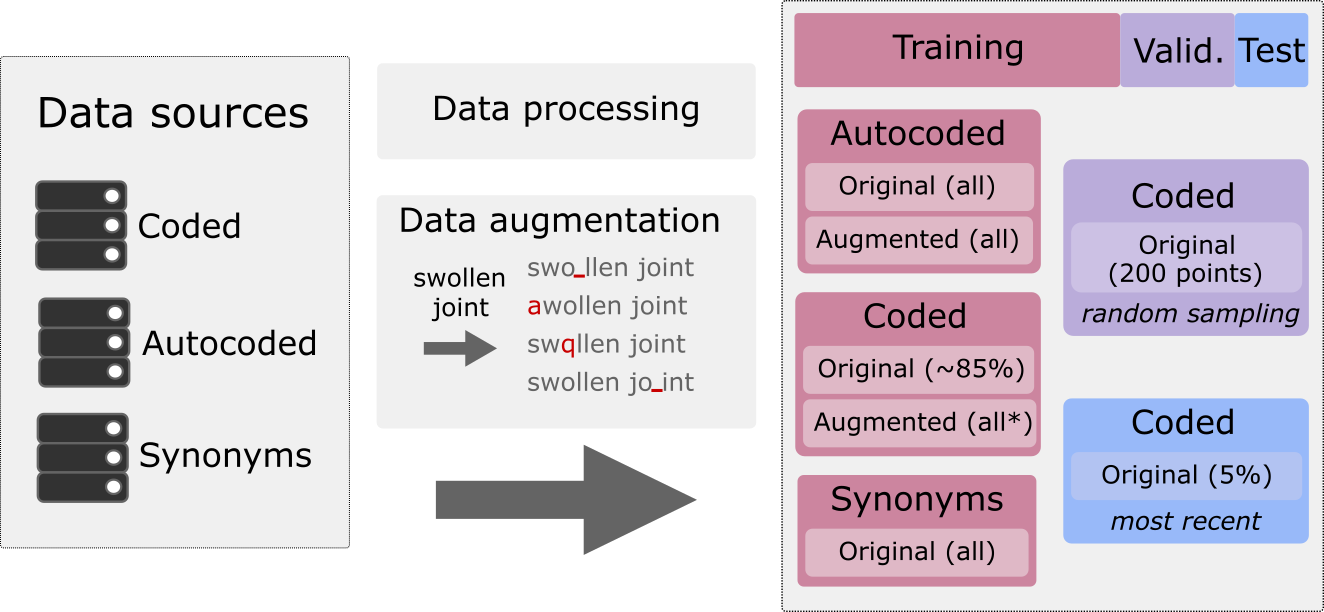}
\caption{From raw data sources to training, validation, and test data via data processing and data augmentation (\textsc{TARS} and \textsc{xTARS} models). Data coming from the ontology is omitted for computational reasons.}
\label{fig:data-split-xtars}
\end{figure}

For the \textsc{xTARS} experiments, we train only on coded, autocoded, and synonyms (including augmented data). We omit the ontology data for computational reasons.
Still, the \textsc{xTARS} model can in principle predict for all LLTs in MedDRA (including LLTs not included in the training set) because the underlying BERT model is able to propose all LLTs from MedDRA, and the \textsc{xTARS} model is able to make zero-shot predictions.
As validation set, we only take 200 samples (instead of the full validation set) because in the current implementation, all possible classes are passed to \textsc{xTARS} for validation during training, resulting in a computational cost of $\sim 80\,000$ prediction for each validation sample. At prediction time, however, only the top-5 candidate classes from the BERT models are passed to \textsc{xTARS}, as outlined in the main text.

\subsection{Biomedical language models} 

From the BLURB leaderboard \citep{gu-2020-pubmedbert}, we take the best performing language models for biomedical tasks (as of July 2021):

\begin{itemize}

\item \textbf {bioBERT}\citep{lee-2019-biobert} is a BERT model trained on a biomedical corpus via a mixed domain pre-training strategy.
The starting point is a standard BERT model pretrained on general corpus such as Wikipedia and BookCorpus.
From that, the pretraining process is continued using biomedical text, namely PubMed abstracts (PubMed) and PubMedCentral (PMC) full text articles. We use BioBERT v.1.1.

\item \textbf {PubMedBERT}\citep{gu-2020-pubmedbert} is a BERT model which - in constrast with bioBERT - is pretrained exclusively on biomedical text; specifically, it does not use the BERT weights as initialization, and it builds the vocabulary from scratch based on the biomedical text. The training corpus is also PubMed and PMC. A larger batch size w.r.t. bioBERT (8,192 vs 192) is used in the pretraining process.

\item \textbf {sciBERT}\citep{beltagy-etal-2019-scibert} is a BERT model which is also pretrained from scratch. Differently from PubMedBERT (and bioBERT), sciBERT is trained on a corpus comprising both computer science (18\%) and biomedical papers (82\%) from Semantic Scholar \citep{ammar-etal-2018-construction}.

\end{itemize}

\subsection{Details on model training}

\subsubsection{BERT models}
Training is performed on the training set  (see Fig. \ref{fig:data-split-bert}) for 20 epochs with Adam optimizer with a learning rate of 1e-4 and batch size of 512. Learning rates of 1e-5 and 5e-5 were also evaluated, but yield a  lower performance. The same hyper-parameters are used for each language model. Training a single model takes approximately 7~h on 4 Tesla V100 GPUs. We select the model with the highest accuracy on the validation (holdout) data, and we evaluate the model on the test data.

\subsubsection{\textsc{TARS} and \textsc{xTARS} models}
Training is performed on the training set (see Fig. \ref{fig:data-split-xtars}) for 5 epochs with Adam optimizer with a learning rate of 5e-5 and batch size of 32. Learning rates of 1e-5, 3e-5, 7e-5, and 8e-5 were also evaluated, but yield a  lower performance. The same hyper-parameters are used for each language model. Training a single model takes approximately 28~h on 1 Tesla V100 GPUs. We select the model with the highest accuracy on a subset of 200 samples of validation (holdout) data due to computational cost. We evaluate the model on the test data.

\subsection{Results on the test set, split by class frequency}

Results split by class frequency are presented in Table \ref{llt-benchmark-table-frequency}.
\begin{table*}[h!t]
\small
\centering
\begin{tabular}{l X l X l X l X l X l X l X l X l X l }
\hline
\hline
\multirow{2}{4em}{Model} &  
\multicolumn{8}{| c|}{LLT Accuracy [\%]} \\
& All & $k$=0 & $k$=1 & $k$=2 & $k$=3 & $k$=5 & $k$=10 & $k$>=100 \\

\hline
PubMedBERT (PMB) (single)  & 74.9$_{0.5}$ & 31.0$_{1.8}$& 44.5$_{0.8}$ & 57.1$_{1.3}$ &  60.3$_{1.8}$ & 68.5$_{1.7}$ & \textbf{65.6}$_{1.2}$ & 84.0$_{0.7}$  \\
BioBERT (BB) (single)   & 74.9$_{0.1}$ & 29.2$_{1.9}$& 41.8$_{1.5}$ & 53.5$_{1.5}$ &  60.0$_{2.6}$ & 65.9$_{2.6}$ & 62.5$_{1.1}$ & 84.4$_{0.4}$  \\
sciBERT (SB) (single)   & 74.9$_{0.3}$ & 27.8$_{1.7}$& 42.5$_{2.5}$ & 58.4$_{1.3}$ &  57.5$_{2.2}$ & 66.4$_{1.1}$ & 60.3$_{2.4}$ & 84.6$_{0.8}$  \\
\hdashline[0.5pt/5pt]
TARS (neg=2 cos)  & 64.9$_{1.2}$ &  \textbf{44.0}$_{1.2}$ & \textbf{52.3}$_{1.9}$ & 58.2$_{1.2}$ & 56.1$_{1.3}$ & 55.0$_{1.9}$ & 58.5$_{1.7}$ & 69.8$_{3.0}$  \\
TARS (neg=10 cos) & 62.6$_{0.8}$ & 39.8$_{0.4}$ & 47.4$_{1.0}$ & 53.8$_{4.6}$ & 56.3$_{3.9}$ & 52.8$_{1.7}$ & 52.7$_{2.8}$ & 66.4$_{1.3}$ \\
\hdashline[0.5pt/5pt]
\textsc{xTARS} (neg=10 cos) & 64.9$_{0.9}$ & 42.2$_{2.3}$ & 51.0$_{1.2}$ & 56.4$_{0.7}$ & 56.7$_{2.5}$ & 55.6$_{1.2}$ & 56.0$_{3.5}$ & 71.4$_{1.6}$  \\
\textsc{xTARS} (neg=top-5) & 76.2$_{0.4}$ & 37.0$_{1.1}$ & 48.7$_{1.0}$ & 57.7$_{1.2}$ & 64.6$_{2.2}$ & 67.8$_{1.7}$ & 63.6$_{2.2}$ & 83.9$_{0.7}$  \\ 
\textsc{xTARS} (neg=top-5+5 cos)  & 77.3$_{0.2}$ & 37.9$_{0.2}$ & 47.4$_{3.1}$ & \textbf{59.3}$_{2.0}$ & \textbf{67.4}$_{2.0}$ & \textbf{68.9}$_{2.5}$ & 63.6$_{1.7}$ & 84.8$_{0.3}$  \\ 
\textsc{xTARS} (neg=top-5+10 cos)  & 76.9$_{0.7}$ & 38.5$_{1.1}$ & 48.8$_{0.4}$ & \textbf{59.3}$_{0.5}$ & 66.1$_{0.8}$ & 68.8$_{1.1}$ & 63.3$_{1.1}$ & 84.6$_{0.9}$  \\
\textsc{xTARS} (neg=top-5+5 cos; $T$=1)   & \textbf{77.5}$_{0.3}$ & 37.3$_{0.9}$ & 48.1$_{0.2}$ & 59.7$_{0.9}$ & 64.8$_{1.1}$ & 67.8$_{2.4}$ & \textbf{65.6}$_{0.7}$ & \textbf{84.9}$_{0.3}$  \\
\hline
PMB (3 models) & 77.8 & \textbf{33.0} & 46.0 & 60.4 &63.4& \textbf{70.9}& \textbf{68.9}& 86.6 \\
BB (3 models)& \textbf{77.9} & \textbf{33.0}&  \textbf{47.9} &\textbf{64.3} &\textbf{65.8} & 69.2 & 65.6 & 87.0 \\
SB (3 models) & \textbf{77.9}& 30.8& 43.7& 57.1& 65.2&69.8& 64.7&  \textbf{87.1} \\
\hline
PMB+SB+BB (3$\times$3) & 79.7& 34.8&  49.8& 63.2& 63.4& \textbf{73.8}& \textbf{70.6}&  \textbf{88.2} \\
\textsc{xTARS} (PMB+SB+BB, 3$\times$3) &\textbf{80.4} & \textbf{48.9}&\textbf{52.6}&  \textbf{64.8} & \textbf{68.3} & 71.5& 68.9& 86.9 \\
\hline
PMB+SB+BB (3$\times$5) & 80.1& 36.2& 48.8& 64.8& 67.1& 73.8& 68.9& 88.5\\
\hline
\hline
\end{tabular}
\caption{\label{llt-benchmark-table-frequency} LLT accuracy on the test set. For details on the models, please see Table \ref{benchmark-table} in the main text. $k$ refers to the number of training samples, excluding augmented data. In this setting, BERT models can perform zero-shot ($k$=0) predictions because the training set contains augmented samples from the ontology for all categories (see Sec. \ref{sec:data-pp}), even when no real samples are present in the training data (i.e. $k$=0).}
\end{table*}

\subsection{Machine learning solution architecture} 

\begin{figure}[htb]
\centering
\includegraphics[width=0.45\textwidth]{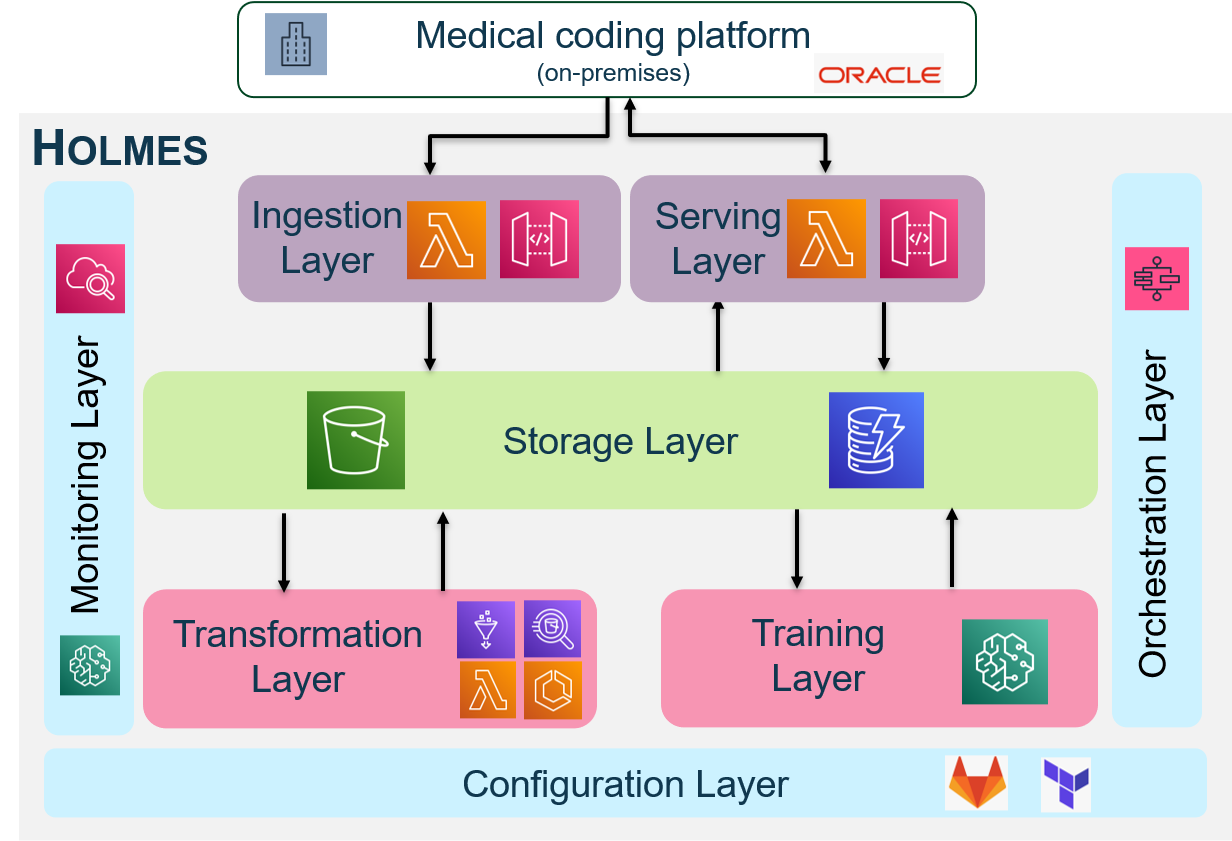}
\caption{Cloud architecture of the deployed medical coding system outlined in the main text.}
\label{fig:architecture}
\end{figure}

Designing and implementing machine learning systems is challenging since they exhibit a different behavior than traditional software systems \citep{sculley2015}.
The cloud architecture of the medical coding (MC) system outlined in the main text (termed Holmes) is organized in separate yet interconnected layers.

\paragraph{Serving layer.} It is responsible for receiving RTs from the MC platform, forwarding them to Holmes, and returning the respective predictions. 

\paragraph{Ingestion layer.} Via the ingestion layer, the MC platform sends all available data needed for model training to Holmes.

\paragraph{Storage layer.} In the storage layer we save all data, solutions, and model versions that are received or created by Holmes. It can be considered as the hard-drive of Holmes.

\paragraph{Transformation layer.} It handles all steps required to make the data received from the MC platform ready for model training. It implements all pre-processing steps, including data augmentation and split into training, validation, and test data.

\paragraph{Training layer.} It performs model training with data transformed by the transformation layer, and saves the trained models to the storage layer.

\paragraph{Configuration layer.} The entire architecture is defined, configured, and created by the configuration layer via infrastructure as code. This allows to install the entire infrastructure via simple scripts.

\paragraph{Monitoring layer.} It observes the system and raises alerts if unusual behavior is detected, e.g. requests from unknown IP addresses.

\paragraph{Orchestration layer.} It schedules all tasks such as model training or data processing in the right order.

\subsection{Graphical user interface of the medical coding platform}

\begin{figure*}[htb]
   \centering
    \includegraphics[width=\textwidth]{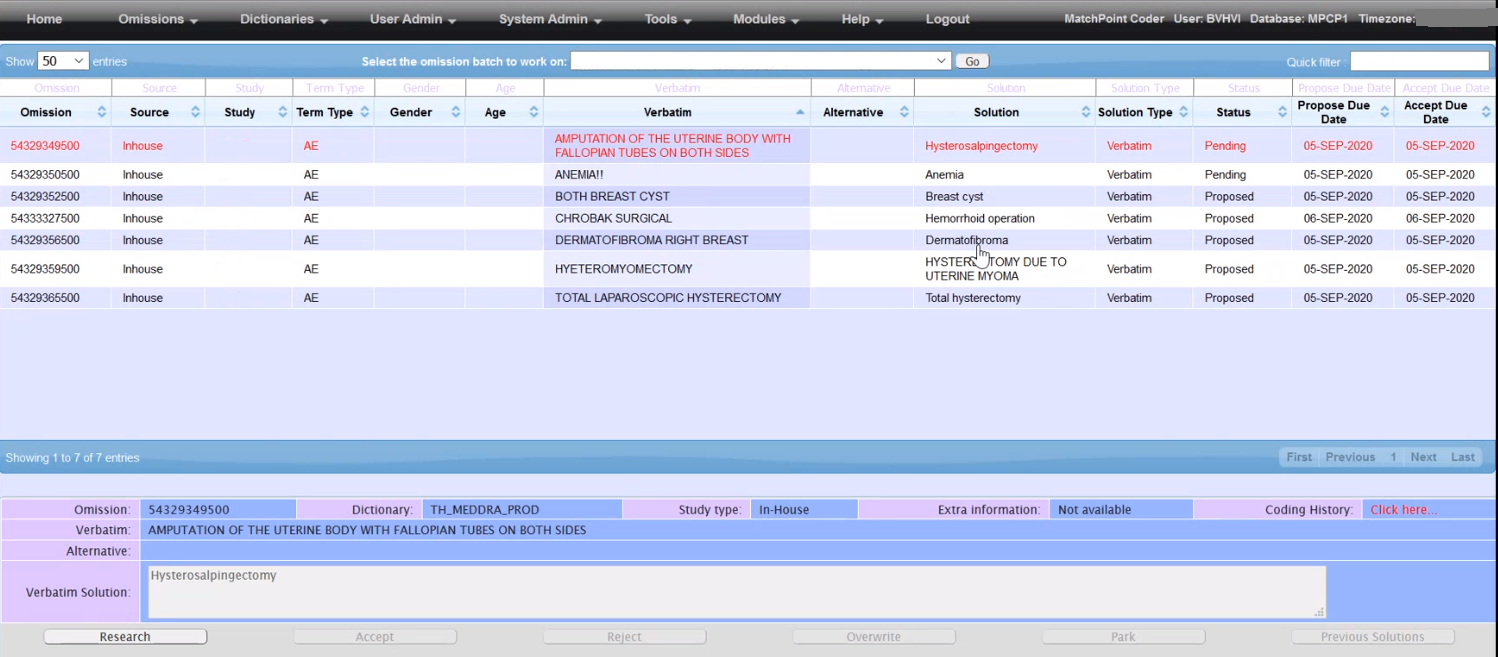}
\caption{Screenshot of the medical coding platform where the coding solutions proposed by the algorithm described in the main text are shown to medical coders for acceptance or rejection.}
\label{fig:mpc-screenshot}
\end{figure*}

\end{document}